\documentclass{aa}
\usepackage{graphicx}
\def\ssim{\setbox0=\hbox{$\sim$}%
\setbox1=\hbox{$<$}\dimen0=\ht1%
\advance\dimen0by-1.2pt\,\lower.6\dimen0%
\copy0\kern-\wd0\raise.4\dimen0\copy1 \,}

\def\gsim{\setbox0=\hbox{$\sim$}%
\setbox1=\hbox{$>$}\dimen0=\ht1%
\advance\dimen0by-1.2pt\,\lower.6\dimen0%
\copy0\kern-\wd0\raise.4\dimen0\copy1\,}

\def\lambdab{\lambda\mkern-9mu\lower1.2pt\hbox{$\mathchar'26$}}%

\begin{document}
   \title{The origin of primary nitrogen in galaxies}

 \author{G. Meynet
          \&
           A. Maeder
          }

   \offprints{G. Meynet}

   \institute{Geneva Observatory CH--1290 Sauverny, Switzerland\\ email: Georges.Meynet@obs.unige.ch
             }

   \date{Received / Accepted }

\abstract{
We investigate the role of stellar axial rotation on 
the nitrogen nucleosynthesis at low metallicities Z. For this purpose,
we have calculated models with initial masses between 2 and 60 M$_\odot$ at Z=0.00001  
from the zero age sequence to the phase of thermal pulses for models below or equal to 7 M$_\odot$, 
and up to the end of central C--burning for the more massive stars. 
The models include all the main physical 
effects of rotation. We show that intermediate mass stars with rotation naturally 
reproduce the occurrence and amount of primary 
nitrogen in the early  star  generations in the Universe. We identify two reasons why rotating models
at low Z produce primary $^{14}$N: 1) Since the stars lose less angular momentum, they rotate faster. Simultaneously, 
they are more compact, thus differential rotation and shear mixing are stronger.
2) The H--burning shell has a much higher temperature and is thus closer to the core, which favours mixing between the two.
\keywords Physical data and processes: nucleosynthesis -- Stars: evolution -- Stars: rotation
}

   \maketitle
%

\section{The context: observations and models}

For more than 20 years, the  nucleosynthetic origin of primary nitrogen, 
which is one of the most abundant and important elements in the Universe, for life in particular, 
has remained a deep mystery in astrophysics (Edmunds and Pagel \cite{edm}; Barbuy \cite{bar}; Carbon et al. \cite{car}; 
Thuan et al. \cite{thu}; Izotov \& Thuan \cite{izo}; Henry et al. \cite{hen0}). 
Nitrogen is mainly produced by the CN cycle of the CNO reactions which catalyze hydrogen burning in stars. In the CN cycle, the reaction $^{14}$N(p,$\gamma$)$^{15}$O which destroys  nitrogen has a very small cross section, which allows $^{14}$N to accumulate with time (e.g. Arnett \cite{arn}). Thus, $^{14}$N  is usually  the daughter element, hence a secondary element, of the CNO  initially present in stars. If nitrogen is secondary, the increase in the abundance of $^{14}$N should be proportional to the initial CNO content and  consequently  the $^{14}$N--content will be proportional to the square of the CNO and metal content in a galaxy. If  $^{14}$N comes from a primary production, {\it i.e.} is produced from the original hydrogen and helium, the $^{14}$N--abundance is proportional to that of the  other primary heavy elements (Talbot \& Arnett \cite{tal}). 

The first evidence of a primary production of $^{14}$N in the early phases of the evolution of our Galaxy came from the study of the very old and low metallicity stars. Several authors (Edmunds \& Pagel \cite{edm}; Barbuy \cite{bar}; Tomkin \& Lambert \cite{tom}; Matteucci \cite{mat}; Carbon et al. \cite{car}; Henry et al. \cite{hen0}) have shown that the ratio N/O of  nitrogen to oxygen remains constant with a plateau at log N/O $\simeq$ -1.6  in the early evolution of the Galaxy, thus implying a primary origin of nitrogen. Later when the metal abundance has reached  about 1/5 of the solar abundance, the N/C and N/O ratios grow rapidly, as expected for secondary elements. Thus, primary production seems to occur only in the very early phases of galactic evolution, while  secondary production appears later, when the metal content of a galaxy is higher. A second piece of compelling evidence is provided by the study of the  N/O ratios in ionized HII regions of blue compact dwarf galaxies (Thuan et al. \cite{thu}; 
Izotov \& Thuan \cite{izo}; Izotov \& Thuan \cite{iz0}). There also  N/O is observed to be constant at very low metallicities Z. The blue compact galaxies are in general galaxies where the  average star formation rate has been very low in the past and which are therefore still in an early stage of their chemical evolution. An example is the galaxy IZw~18, which has the lowest known metallicity (1/50 of solar), and which shows indications of primary nitrogen (Kunth et al. \cite{kun}; Izotov \& Thuan \cite{izo}).

A third indication for  the presence of primary nitrogen comes from the observed N/O gradient in spiral galaxies. If nitrogen is a purely secondary element, the N/O gradient should be equal to the gradient of heavy elements, and in particular to the  gradient of O/H in these galaxies. The fact that the N/O gradient is  relatively flat at low metallicity Z  indicates that production of nitrogen is dominated by primary processes at low Z and secondary processes at high Z (Garnett et al. \cite{ga97}; Ferguson et al. \cite{fer}; Henry \& Worthey \cite{hen}). A problem has arisen  some years ago, because some low Z damped Ly$\alpha$ systems, which are likely protogalactic structures at relatively high redshifts, have N/O ratios lower than those observed in the  HII regions  of blue compact dwarf galaxies of the same Z (Pettini et al. \cite{pet}). However, the apparent discrepancy  has been resolved  by models of damped Ly$\alpha$  systems which account for both ionized  and neutral regions (Izotov et al. \cite{isc}).

Observationally, it is still uncertain whether the observed primary nitrogen originates from massive or from intermediate mass stars. The claims that primary nitrogen is made in massive stars are mainly based on the low scatter of the observed N/O ratios at low Z
(Matteucci \cite{mat}; Thuan et al. \cite{thu}; 
Izotov \& Thuan \cite{izo}; Izotov \& Thuan \cite{iz0}). The  argument is that, if nitrogen is synthesized in massive stars, there is no time delay between the injection of nitrogen and oxygen and thus a rather small scatter would result. On the contrary, if the primary nitrogen is made in intermediate mass stars, the N/O ratio increases with time, since these stars release their  nitrogen much later than the oxygen made in massive stars. This would lead to a larger scatter in the observations of galaxies at various stages of their evolution. Some recent studies found that a  significant scatter does exist (Garnett \cite{gar}; Skillman et al. \cite{ski}). Also, Henry et al. (\cite{hen0}) have calculated models which  support the view that intermediate mass stars between 4 and 8 M$_\odot$, with an age of about 250 Myr, are likely to dominate the nitrogen production.

Let us examine  what stellar models tell us. The conditions needed for the  production of primary nitrogen are very simple. In a star which has both a helium burning core and a hydrogen burning shell, some amount of the new  carbon synthesized in the core must be transported into  the hydrogen burning shell, where the CNO cycle will convert it into  primary $^{14}$N . For massive stars, only models with ad hoc hypotheses are able to do this. For example, in some models (Timmes et al. \cite{tim}), mixing has been arbitrarily ``permitted'' between the helium-- and hydrogen--burning zones. Without any physical explanation, it is difficult to understand why the production of primary nitrogen only occurs at low metallicities. Models of metal free Population III stars (Umeda et al. \cite{ume}) produce some primary nitrogen, but in too low quantities. Up to the  phase of thermal pulses, asymptotic giant branch (AGB) stars produce no primary nitrogen. Only during thermal pulses, some helium products might be transported into the hydrogen burning shell to produce primary nitrogen. However, the existing AGB models
(Renzini \& Voli \cite{ren}; Marigo \cite{mar}) are ``synthetic'' models, which means that the model parameters follow some analytical relations that have been fitted to the observations. While this may be useful for some purposes, it cannot be claimed that it represents consistent physics  leading to primary nitrogen production.

\section{Primary nitrogen production in rotating models}

We construct stellar  models with axial rotation and  very low metallicity Z, typical of the early star generations. Rotation is clearly important, because there are indications from studies in the Magellanic Clouds that stars may on the average rotate faster at lower Z (Maeder et al. \cite{mgm}). Our models include the main effects of rotation on stellar evolution: the hydrostatic effects, the meridional circulation which transports both angular momentum and chemical elements (Maeder \& Zahn \cite{maz}),  the effects of horizontal turbulence (Zahn \cite{zah}),  the effects of shear mixing
(Zahn \cite{zah}; Maeder \& Meynet \cite{MMVII}) and also the effects of rotation on the mass loss rates (Maeder \& Meynet \cite{MMVI}). Such model physics has already been applied to stars of solar composition and to the Magellanic Clouds. In both cases these models have resolved major discrepancies (Meynet \& Maeder \cite{MMV}; Maeder \& Meynet \cite{MMVII}). 
 
 \begin{figure}
\resizebox{\hsize}{!}{\includegraphics{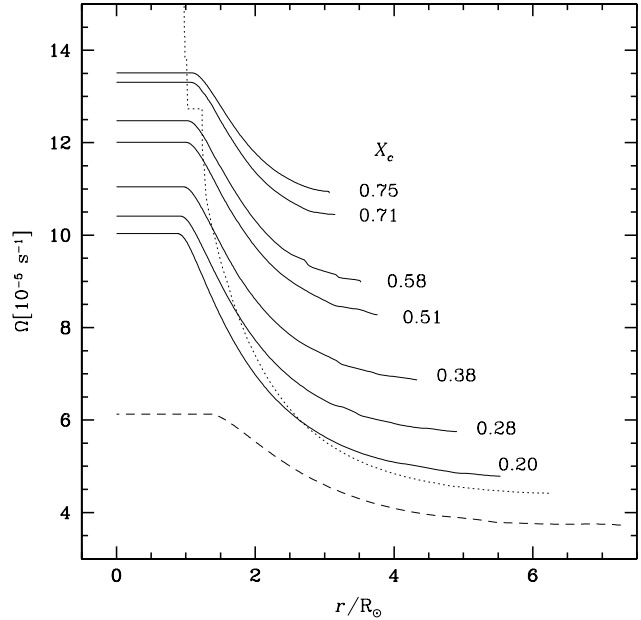}}
\caption{Evolution of the distribution of angular velocity $\Omega$  as a function of the distance to the center in a 20 M$_\odot$ star with an average velocity of 230 km s$^{-1}$ and Z=0.00001. $X_{\rm c}$   is the hydrogen mass fraction at the center. The dotted line shows the profile when $X_{\rm c}$=0. The dashed  line shows the same at Z=0.004 when  $X_{\rm c}$=0.28.}
\label{fom}
\end{figure}

 We have calculated models of 2, 3, 5, 7, 9, 15, 20, 40 and 60 M$_\odot$ at Z =0.00001  from the zero age sequence to the phase of  thermal pulses for models below or equal to 7 M$_\odot$, and up to the end of central C--burning for the more massive stars. Fig.~1 shows an example of the evolution of  angular velocity $\Omega$ until the end of the Main Sequence phase. The interesting property of very low Z models is that the $\Omega$--gradients are much steeper than at higher Z, because of  the lower mass loss rates at lower Z and  the greater compactness of the star.  The consequence of the larger $\Omega$--gradients is that there is more mixing due to the shear instability and thus shallower internal molecular weight gradients ($\mu$--gradients). Comparisons of the surface enrichment in A--supergiants in the SMC and in the Milky Way  fully support the presence of more mixing at lower Z (Venn \cite{ven}; Maeder \& Meynet \cite{MMVII}).
 
  \begin{figure}
\resizebox{\hsize}{!}{\includegraphics{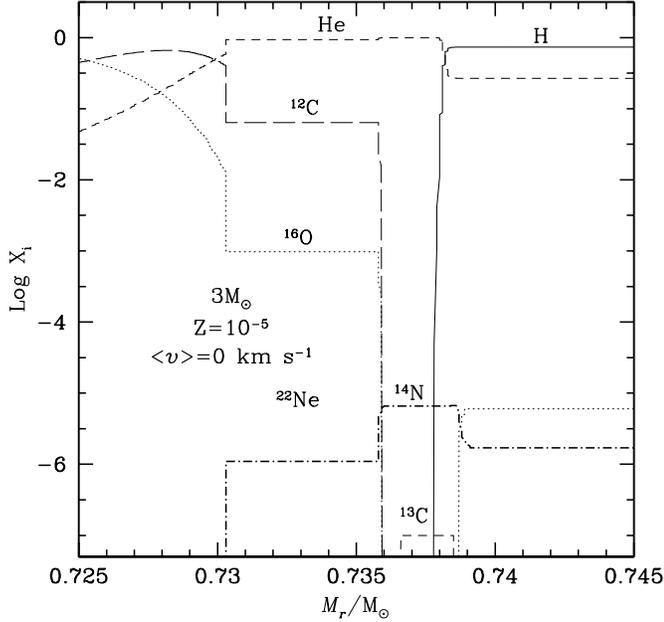}}
\caption{Distribution of the main elements in the region of the helium--  and hydrogen--burning shells during the TP--AGB phase of  an initial non--rotating 3 M$_\odot$ star at Z=0.00001.}
\label{N1430}
\end{figure}

  \begin{figure}
\resizebox{\hsize}{!}{\includegraphics{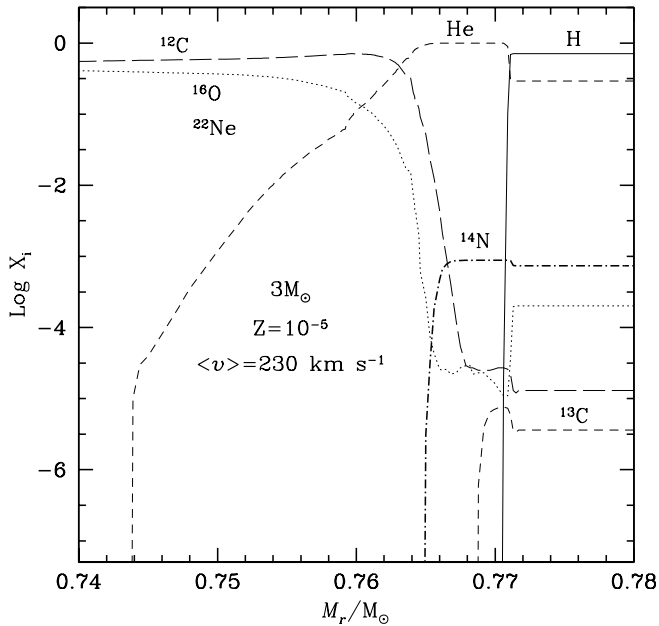}}
\caption{Same as Fig.~\ref{N1430} for a 3 M$_\odot$ rotating model during the TP--AGB phase. The average rotational velocity during the Main Sequence is
230 km s$^{-1}$.}
\label{N1433}
\end{figure}

The combination of the effects of low Z and rotation enables primary nitrogen production 
during the core He--burning phase, in the zone between the core and the H--burning shell. 
Later, on the AGB,
the zone between the He-- and the H--burning shells, enriched in primary nitrogen, shrinks, but at the same time more and more primary $^{14}$N is 
entering the outer convective zone. This primary $^{14}$N in the outer envelope will escape from the star, either in the  stellar wind or during  the formation of a planetary nebula.
This mode
of nucleosynthesis is quite different from the classical scenarios in which the primary nitrogen is produced
during the thermal pulses (TP) AGB phase.
Fig.~2 and 3 show the abundances of some nuclei in the intershell region in  TP--AGB models  of 3~M$_\odot$
models, with and without rotation, at the luminosity of the 5th thermal pulse.  We notice first that the CO core is slightly larger with rotation and that the chemical gradients at its outer edge are also shallower. The most  remarkable difference is the much higher $^{14}$N--content in the intershell region (i.e. where helium has its highest plateau) and in the adjacent external deep convective zone. This $^{14}$N--content is orders of magnitude higher than the total abundance of original CNO elements and it is thus of primary origin.   
The models between 2 and 7 M$_\odot$ show a similar behaviour. 
We emphasize also that similar models with rotation at Z=0.020 do not produce any primary nitrogen (Meynet \& Maeder \cite{MMV}). 
The mechanism which produces primary nitrogen only works efficiently for Z less than $\sim 1/5$ of solar.

There are two physical reasons for this production of primary $^{14}$N in rotating models at low Z . 1) As explained above, the $\Omega$--gradients are steeper, while the $\mu$--gradients are shallower at lower Z.  These effects are present during the whole evolution and  enhance shear mixing. Thus, at very low Z with rotation, there is a very efficient transport of new $^{12}$C  from the core  to the H--burning shell, where the CNO cycle turns it to primary  $^{14}$N. 2) The second reason is that at lower Z  the CNO burning occurs at much higher temperature $T$.  Indeed,  the amount of energy  produced  in TP--AGB stars is of the same order of magnitude, whatever the value
of Z.  In order to maintain the rate of energy production when, as in the present models, the CNO content is about 2 $\cdot$ 10$^3$ times smaller than solar, $T$ has to be about  ($2 \cdot 10^3)^{(1/9)} = 2.3$ times higher, because the energy production of CNO  is proportional to $T^9$ at the typical  $T = 8 \cdot 10^7$  K, where H--shell burning occurs in TP-AGB stars of  very low Z. Since the H--burning shell needs to be at much higher  $T$ in lower Z stars, it is much closer to the edge of the He--burning shell, which is sitting on the outer edge of the CO core and  which produces $^{12}$C. The value of $T$ in the CO--core is not very sensitive to the initial composition.  Thus, the intershell distance is much shorter at very low Z  (by about an order of magnitude in the present case) and the transport of $^{12}$C from one shell to the other  is easier. 
 
 \section{The nitrogen stellar yields at low metallicity}
 
   \begin{figure}
\resizebox{\hsize}{!}{\includegraphics{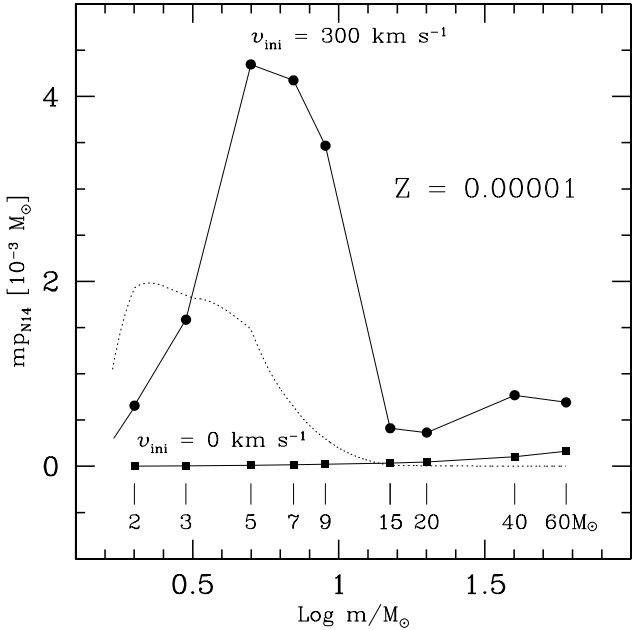}}
\caption{Amount of nitrogen (mainly primary in rotating models)
produced as a function of the initial stellar mass. The quantities given here are the so--called  stellar yields ${\rm mp_{N14}}$, where m is the initial mass and  ${\rm p_{N14}}$  its fraction in form of total ejected $^{14}$N. The initial rotation velocity $v_{\rm ini}$ on the ZAMS is indicated. For the initial mass
stars between 3 and 15 M$_\odot$, the average rotational velocity during the Main Sequence is around 230 km s$^{-1}$. The yields weighted by the Salpeter mass spectrum
is shown by a dotted line (arbitrary units).}
\label{mpn145}
\end{figure}
 
Fig.~4 shows the amount of nitrogen, the so--called stellar yields  ${\rm mp_{N14}}$ as a function of the initial stellar masses. We notice that intermediate stellar masses of 5 to 7 M$_\odot$ have the largest stellar yields of primary nitrogen. When these yields are weighted by the Salpeter IMF (Fig.~4), we see that the main contributors are
between 2 and 5 M$_\odot$.
For the initial stages of the evolution of galaxies considered here, we may to first approximation take the ratio of the net yields  in nitrogen and oxygen as an estimate for the observed N/O ratio. For the production of oxygen, we use the relation between the mass $M_\alpha$  of the He--C--O core at the end of the He or
C--burning phase and the yield in oxygen by Arnett (\cite{ar9}).  Taking both the N and O productions for the models at Z = 0.00001 and convoluting them with a standard initial mass function, we get  a ratio log N/O = -1.88. This result is within a factor of two of the value log N/O $\simeq$ -1.6 observed at low Z.  This is a very satisfactory  agreement for two reasons. 1). Here, we have considered models with an average velocity of 230 km s$^{-1}$. However, the production of primary $^{14}$N increases rapidly with rotation, thus the production for the average velocity is smaller than the average production for the actual distribution of rotational velocities. Moreover, there seems to be more fast rotators at lower Z
(Maeder et al. \cite{mgm}).  2). Nitrogen is ejected mainly by AGB stars with ejection velocities of a few 100 km s$^{-1}$, while oxygen is ejected by supernovae  at much higher velocities of $10^4$  km s$^{-1}$ or more. Thus, a fraction of the oxygen produced may escape from the parent galaxy, leading to a higher N/O ratio than in the simple estimate made here. 

\section{Conclusion}

For the first time, we present 
models of star evolution with rotation and very low 
metallicities, typical of the first stellar generations 
in the Universe.
We show that stellar rotation in very low Z
models leads to the production of primary nitrogen 
in amounts that are in global agreement with those observed.
The main sources are intermediate mass stars with initial 
masses between 2 and 5 M$_\odot$ (see Fig.~4). 
The yields of s--process elements at low metallicity will also be deeply affected by rotation (Langer et al. \cite{lhw}).
Indeed, as can be seen from Figs.~2 and 3, the abundance of the main neutron source $^{13}$C
is considerably enhanced in the intershell region of the rotating model. Finally, we emphasize
that a certain amount of primary $^{22}$Ne should also be synthesized as a result of the burning of primary $^{14}$N
in the He--burning shell. However, contrary to primary $^{14}$N, the exact yields in $^{22}$Ne are difficult to predict, since
they critically depend on the mass loss and mass cut leading to the final remnants.

\end{document}